\documentclass[a4paper]{jpconf}
\usepackage{graphicx}

\usepackage{amsmath}
\usepackage{cite}
\usepackage[T2A]{fontenc}
\usepackage[utf8x]{inputenc} 
\usepackage[%
	breaklinks=true,
	colorlinks=true,
	linkcolor=blue,
	urlcolor=blue,
	citecolor=blue
]{hyperref}
 
\begin{document}
\title{Monte Carlo simulation of colloidal particles dynamics in a drying drop}

\author{K.\,S.~Kolegov}

\address{Volga State University of Water Transport, Caspian Institute of Maritime and River Transport, 2 Kostina St., Astrakhan, 414014, Russia}

\address{Astrakhan State University,  20A Tatishchev St., Astrakhan, 414056, Russia}

\address{Landau Institute for Theoretical Physics Russian Academy of Sciences, 1-A  Academician Semenov Ave., Chernogolovka, 142432, Russia}

\ead{konstantin.kolegov@asu.edu.ru}

\begin{abstract}
The work is devoted to particles dynamics simulation in a colloidal drop, when it dries on a substrate and the triple-phase boundary is fixed. Experimental observations [Deegan~R.\,D. et. al., 2000] show a ring deposition on a solid substrate after full droplet desiccation. This phenomenon of macrolevel is known as coffee ring effect. There are other experiments that show a co-effect at the microlevel. We are talking about the formation of a quasicrystalline structure on the outer part of the ring and occurrence amorphous inner zone of ring [Mar\'{\i}n~\'A.\,G. et. al., 2011]. The goal of this work is to check the hypothesis of other authors that this phenomenon is explained by the competition between the characteristic times of diffusional displacement of particles and their transfer by a compensation flow [Mar\'{\i}n~\'A.\,G. et. al., 2011]. Numerical calculations using a model built on the basis of such assumptions and effects did not show the formation of a quasicrystal structure. It is probably necessary to take into account additional effects in such a system, for example, surface tension.
\end{abstract}

\section{Introduction}

The processes of heat and mass transfer in drops drying on substrates are interesting because of their different applications. We list only some of them, for example, evaporation of picoliter droplets under reduced pressure for cooling the surface~\cite{Saverchenko2015}, technology for assessing the quality of drinks~\cite{YAKHNO2018451}, removing nanoparticles from a solid surface~\cite{MAHDI201513}, and so on.

The coffee ring effect is studied theoretically and experimentally  in~\cite{Deegan2000}. Capillary flows transfer colloidal particles to the three-phase (solid--liquid--gas)  contact  line during the process of the droplet drying on the substrate.  The formation of a ring deposition is observed, if the three-phase boundary has been fixed throughout the process (pinning). There are many continual models describing this phenomenon at the macrolevel, for example,~\cite{Deegan2000, Fischer2002, TarasevichModernPhys2011, Ozawa2005} etc. However, this approach does not allow to predict the structure of the particles relative to each other (microlevel). The first few rows of polystyrene particles near the contact line according to the experiment~\cite{Marin2011} form a kind of quasicrystal (particles are close packed). There is a transition to the amorphous zone at a short distance from the contact line,  where the particles are located more or less chaotic. The authors~\cite{Marin2011} explain this phenomenon by the ratio of the characteristic times of diffusion ordering of particles and their transfer by the compensation flow of fluid. This flow is directed to the edge of the droplet where intense evaporation occurs. The velocity of capillary flow in the last about 5\% of the process time increases by more than an order of magnitude~\cite{PhysRevE.83.051602}. The authors~\cite{Marin2011} believe that this rush-hour leads to the fact that the particles do not have time to pack themselves in the process of Brownian motion near the triple line. Thus, there is an amorphous zone of deposition. The authors of another experiment~\cite{Zhang2013}  also study the reasons for the transition from an ordered structure of particles to an unordered one. Their opinion is different, because they believe that the particles at high velocity begin to pile on the formed layers, so there is amorphous deposition.  In addition, the authors~\cite{Zhang2013} suggest that the liquid surface  covering the particles can influence the final form of the structure due to the surface tension when the film rupture. Also, some additional influence may be exerted by the Marangoni flow in the case of a heated substrate~\cite{Zhang2013}. The effect of particle shape on the resulting structure is studied in~\cite{Yunker2011,Yunker2013}.  Understanding this kind of processes is extremely important, for example, in micro- and nanostructuring applications.

There are several discrete models that explicitly describe the dynamics of each individual particle. These models are usually based on Monte Carlo or molecular dynamics methods~\cite{LEBEDEVSTEPANOV2013132,Shi2016}. Some models based on the Monte Carlo method are lattice ones~\cite{ZHANG2016650, Kim2011,jung2014}, others are not~\cite{Petsi2010}.  Often the fluid flow is described continually. Typically, the flow velocity is calculated analytically for simple special cases~\cite{LEBEDEVSTEPANOV2013132,Kim2011,Petsi2010,jung2014,crivoi2014}. The calculated velocity field is superimposed on the particle dynamics. Particle adhesion, wetting and substrate roughness are taken into account in~\cite{LEBEDEVSTEPANOV2013132} .  The simulation results show that the coffee ring effect is the result of compensation flow, low roughness and adhesion. The method used in~\cite{LEBEDEVSTEPANOV2013132} requires a large computing power and is more suitable for a small number of particles in the system. The formation of branched aggregates of nanoparticles in the droplet drying on the substrate is simulated in~\cite{ZHANG2016650}. The models~\cite{Kim2011,crivoi2014} describe the formation of ring depositions during the evaporation of a droplet using a cubic lattice. Cubic particles attached to each other by side faces~\cite{Kim2011,crivoi2014} do not allow to describe the phenomenon of transition from a quasicrystal to an amorphous structure~\cite{Marin2011}.   The paper~\cite{Petsi2010} compares Monte Carlo method and lattice Boltzmann equations.  The semi-cylindrical geometry of the droplet is considered. The authors~\cite{Petsi2010} numerically describe the process of deposition of particles on the substrate as a result of their transfer by the compensation fluid flow and their Brownian motion. A computational experiment is carried out for different contact angles (hydrophilic and hydrophobic substrates) and modes of three-phase boundary (pinning and depinning). Exclusion of particle collisions (their intersection or overlapping) is not described in~\cite{Petsi2010}.  It is a disadvantage of the model in case if this condition is not implemented in the algorithm, because such model does not allow to obtain the form of the structure at the microlevel. At least such results are not presented in~\cite{Petsi2010}. The lattice model is used to mathematically describe the coffee ring effect in~\cite{jung2014}. The circular lattice used in~\cite{jung2014} by definition does not allow to explain the reason for the formation of a quasicrystal at the drop edge, since this structure is already predetermined by the symmetrical arrangement of cells.

The aim of this paper is to determine whether the formation of a structure in which the ordered arrangement of particles (quasicrystal) transforms into an amorphous form is a consequence of convection and diffusion only.

\section{Problem and method description}

\subsection{Physical formulation}

Consider an axisymmetric thin drop with an initial contact angle $\theta \approx \pi / 6$ and the radius of the base $R \approx 0.5$ mm \cite{Marin2011}.  We mean base is the plane of contact of the liquid and the substrate. Three-phase boundary is fixed, $R=\mathrm{const}$ and $\theta=\theta(t)$, where $t$ is process time.  The duration of the evaporation process up to the complete drying of the droplet $t_{max}\approx 1750$~s \cite{Marin2011}. Capillary forces prevail over gravitational forces in a drop of this size, so the geometry of the free surface shape of the liquid corresponds to a spherical segment.

Let us assume that the particles are in suspension. Brownian motion of particles  dominates their  sedimentation due to the small size ones (particle radius $r_p \approx 1$~$\mu$m \cite{Marin2011}), so gravity is not taken into account. The particles are neutrally charged, so the question of interparticle interaction in this work is not affected. We consider the case when the temperature difference of the liquid, substrate and air is insignificant, so thermocapillary flows are excluded from attention~\cite{Barash2009b}.This paper does not address the description of processes when the substrate is heated~\cite{Gatapova2014} or there is some external heat source~\cite{Kolegov2018}. Evaporation is relatively slow in our case, so we assume that the process is isothermal. Let us consider that the dynamic viscosity of the liquid $\eta$ is a constant in the zero approximation. The diffusion coefficient $D$ is also assumed to be a constant.  The diffusion coefficient of particles is determined by Einstein's formula $D = kT/(6\pi \eta r_p)$, where the Boltzmann constant is denoted by $k$, $T$ is the liquid temperature. Substituting the  water viscosity at room temperature, we obtain the value of the diffusion coefficient $D \approx 2\cdot 10^{-13}$ m$^2$/s. 

\subsection{Mathematical model}

It is convenient to use polar ($r, \phi$) and Cartesian coordinates ($x, y$) at the same time in this problem. We are describing the particle dynamics only in the horizontal plane $xy$, since the thickness of the liquid layer in a thin drop is much less than the diameter of its base.  The problem will be solved not in the whole area, but only in the selected sector of the circle due to the axial symmetry. Let us  use  periodic boundary conditions on the side borders of the sector.
 There is also a movable boundary $R_f$  in the plane $xy$ despite that the case with a pinned contact line is considered. Let the radius of fixation $R_f$ is such a boundary, where the particle size and the thickness of the liquid layer are comparable ($2 r_p\approx h$), at that $R_f< R$. Particles cannot reach the $ R $ boundary in a thin drop, since near this boundary the thickness of the liquid layer is small, $h\to 0$. Therefore, the particles will not move further than $R_f$ because behind this boundary $h < 2 r_p$. The force of the surface tension of the fluid will restrain them. Write an expression for the shape of the droplet surface $h(r,t) = \theta (t) \left( R^2 - r^2 \right)/ (2R)$, where the radial coordinate $r = \sqrt{x^2 + y^2}$. Hence, we obtain the dependence of the fixation radius on time
 \begin{equation}\label{eq:fixing_radius}
  R_f (t) = \sqrt{R^2 - \frac{4 r_p R}{\theta (t)}}.
\end {equation}
The height of the droplet during evaporation decreases, in consequence of which the contact angle decreases too. This evolution of contact angle is explain the movement of the boundary $R_f$. Note that most of the time the value of $R_f$ changes slightly. This radius begins to decrease rapidly only at the end of the process. The contact angle $\theta$ decreases linearly over time, $\theta \to 0$ when $t \to t_{max}$~\cite{PhysRevE.83.051602}.

The radial fluid flow velocity, averaged over the height of the drop, is calculated based on the results from the model~\cite{Deegan2000}, 
 \begin{equation}\label{eq:velocity}
 \bar v_r(r,t) = \frac{R}{4(r/R)(t_{max}-t)}\left(  \frac{1}{\sqrt{1-(r/R)^2}} - \left[ 1-(r/R)^2 \right] \right).
\end {equation}
Let us simulate diffusion as a random walk of particles by the Monte Carlo method.

\subsection{Problem solving algorithm}

We take into account  that there must be several modes of starting the calculation module  (<<diffusion + advection>>, <<diffusion>> and <<convection>>) when developing the program. Thus, it will allow to consider two types of mass transfer separately and both together. Let each particle be in one of three states.  The state <<movement>> is characterized by both diffusion and advective transfer. The status <<diffusion only>> implies only one Brownian motion. Particle is stationary if its status is <<stop>>.

First, we uniformly generate random coordinates of particles ($x,y$) in our domain. Second, we set initial values of their status (<<motion>> or <<diffusion only>> depending on the program start mode) for all particles.  Then we go in a cycle on temporary layers with a step $\Delta t = 0.1$~ms to achieve time $t_{max}$. The time step size is choiced on the basis of a series of computational experiments in such a way that the Einstein relation for the mean square of the displacement $\overline{(\Delta l)^2} = 2D t_{max}$ is carried out, where $$\overline{(\Delta l)^2} = \sum_{j=1}^{t_{max}/\Delta t} \frac{1}{N_p} \sum_{i=1}^{N_p}\left( (x_i^j - x_i^{j-1})^2 + (y_i^j - y_i^{j-1})^2 \right),$$
$N_p$ is the number of particles. 

The program recalculate the value of~$R_f$ according to~\eqref{eq:fixing_radius} at each time step and iterate through all the particles. It is depended on status of particle to shift it or not. The flow velocity for the current coordinate $r$ and time $t$ is calculated based on the expression~\eqref{eq:velocity}. We calculate the distance of the particle displacement by the fluid flow in one time step $\Delta t$ by the formula $\Delta t \bar v_r (r,t)$.  The diffusion displacement distance is defined as $\sqrt{2 D \Delta t}$. The value of the polar angle $\phi$ for the diffusion motion of a particle is generated randomly. This step is not executed if the movement is impossible due to the coming overlapping with any other particle. 

Let us use the following rules. First, a particle goes into the state <<stop>> if this one  was in contact with the radius of fixation $R_f$ and had the status <<motion>> or <<diffusion only>>. A particle cannot move beyond $R_f$ because it is prevented by the surface tension of the liquid. Second, if the moving particle could potentially intersect with the stationary one, it changes the status <<motion>> to the value <<diffusion only>>. Third, if a diffusion particle has moved away from the fixation radius by a distance of more than 10$r_p$, then it becomes mobile again (status <<motion>>). The last condition is valid for the case when the program is started in the <<diffusion + convection>> mode. The program algorithm is written with C++ language.

\subsection{Numerical results}

\begin{figure}[h]
\begin{minipage}{14pc}
\includegraphics[width=14pc]{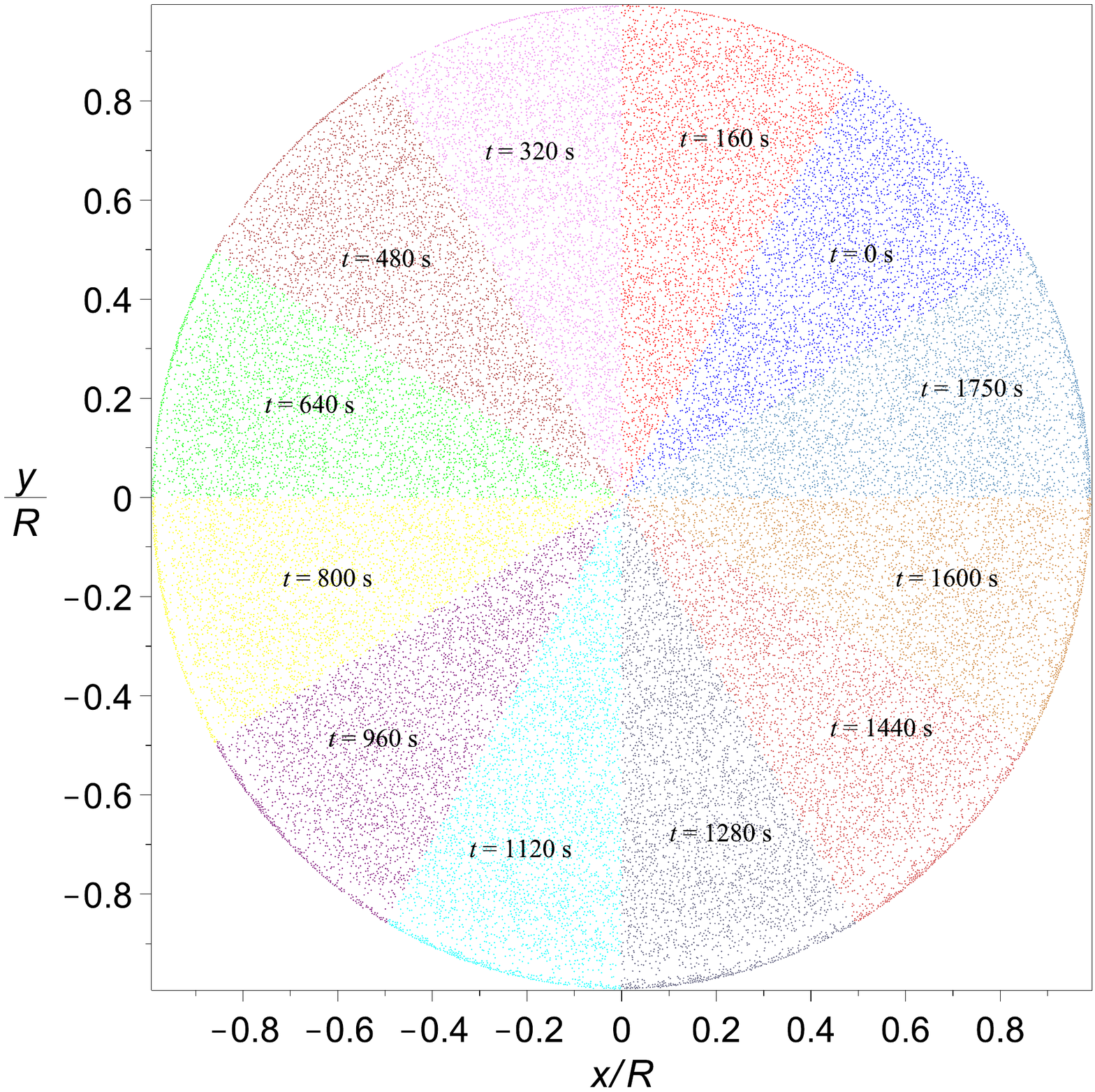}(a)
\end{minipage}\hspace{2pc}%
\begin{minipage}{14pc}
\includegraphics[width=14pc]{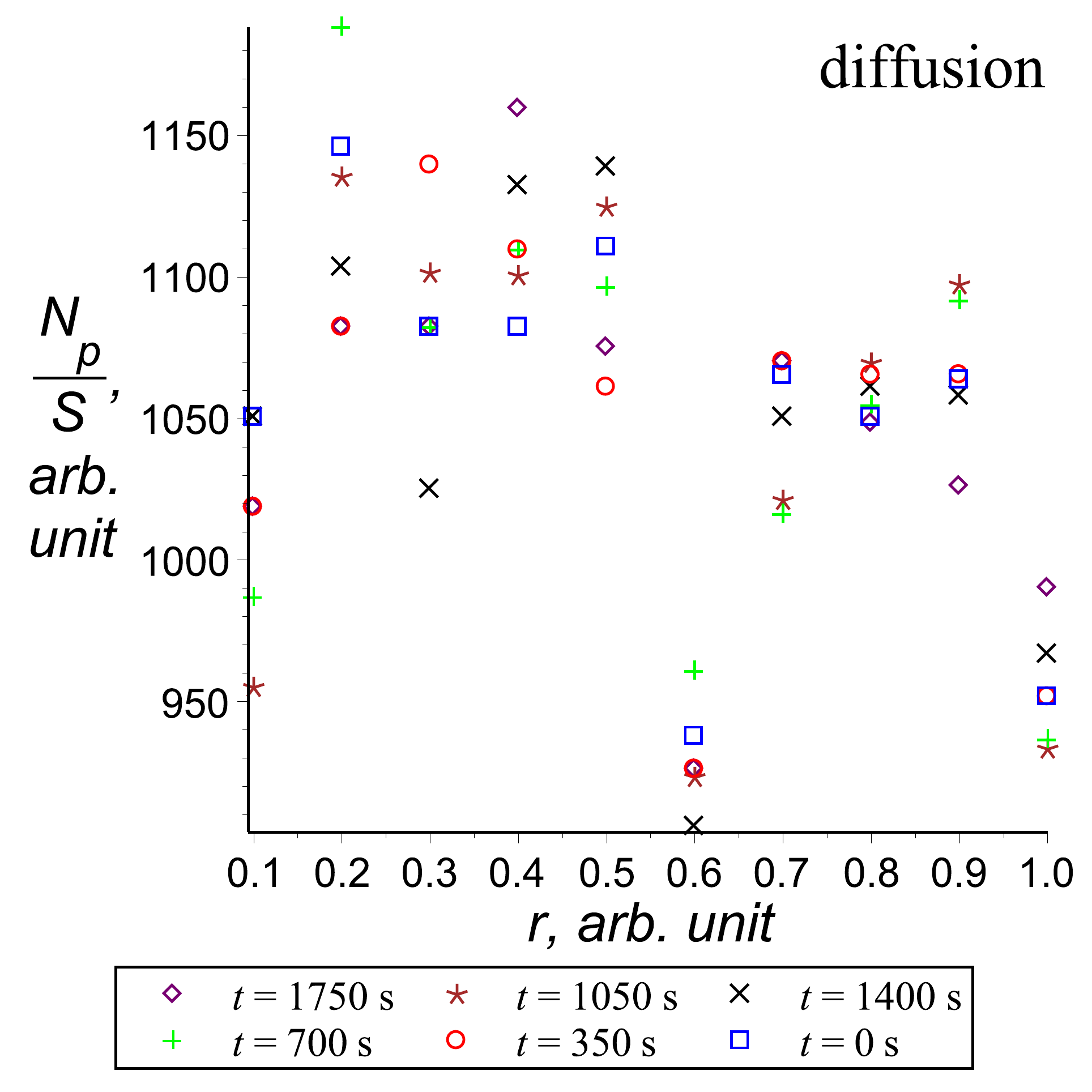}(b)
\end{minipage} 
\caption{Dynamics of colloidal particles in the case of diffusion motion: (a) arrangement of particles at different times, (b) the density distribution of particles.}
\label{fig:diffusion_structures_and_densities}
\end{figure}

The fig.~\ref{fig:diffusion_structures_and_densities} shows the simulation results of the case when the particles only diffuse. This is possible with a uniform evaporation of the liquid along the free surface of the drop~\cite{Deegan2000}. Therefore, compensation flows are absent. The arrangement of the particles in several consecutive moments of time is shown in the fig.~\ref{fig:diffusion_structures_and_densities}a. It can be seen that the final structure ($ t = $ 1750 s) is characterized by a more or less uniform distribution. In addition, in this structure we observe a thin layer of particles near the contact line. These particles hit the fixation radius as a result of chaotic Brownian motion, when the $ R_f $ boundary moved very slowly. The number of particles per unit area (density distribution) is shown in the fig.~\ref{fig:diffusion_structures_and_densities}b. The density of the particle distribution is approximately $1050 \pm 150$ at all considered time intervals.

\begin{figure}[h]
\begin{minipage}{14pc}
\includegraphics[width=14pc]{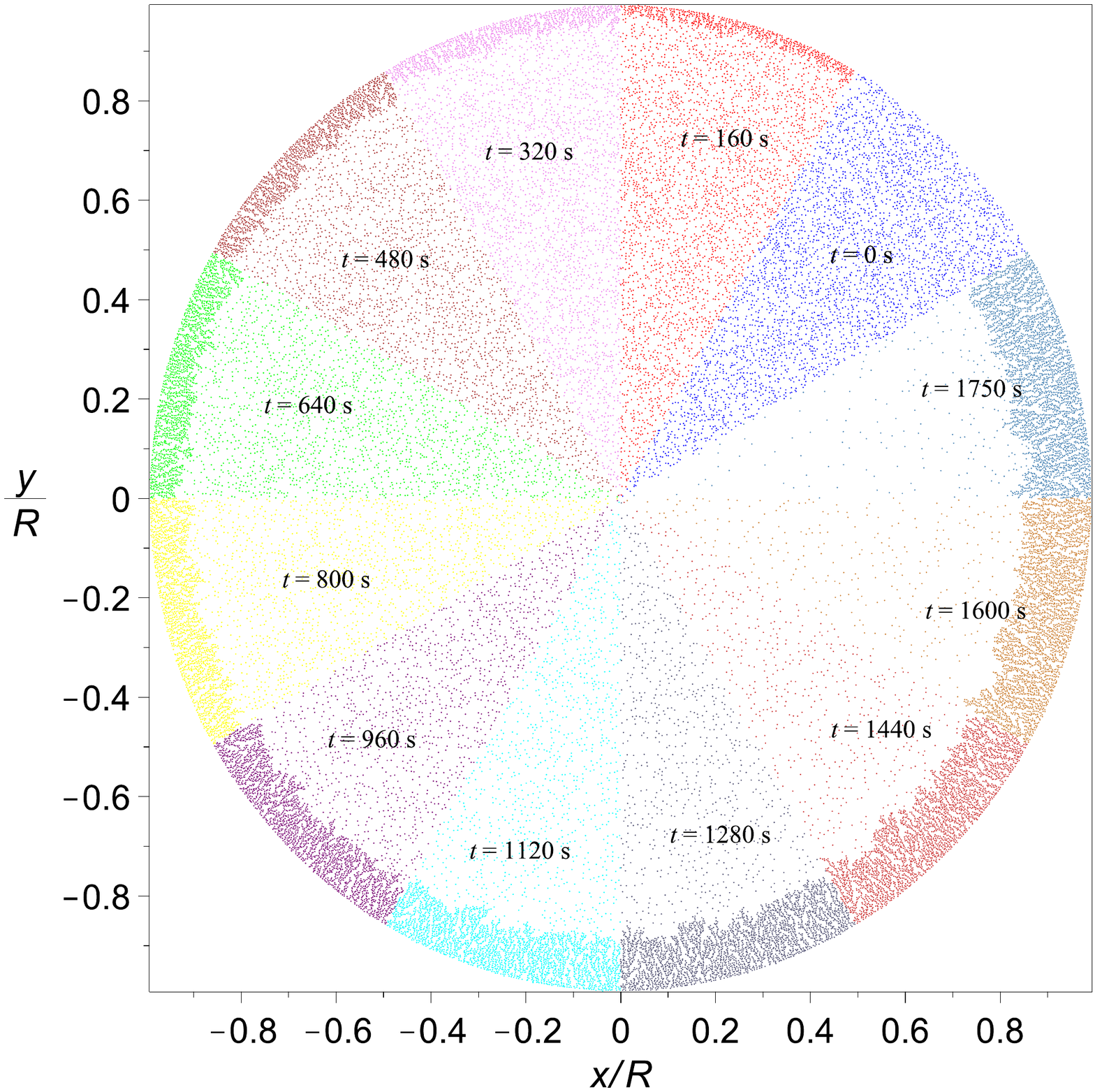}(a)
\end{minipage}\hspace{2pc}%
\begin{minipage}{14pc}
\includegraphics[width=14pc]{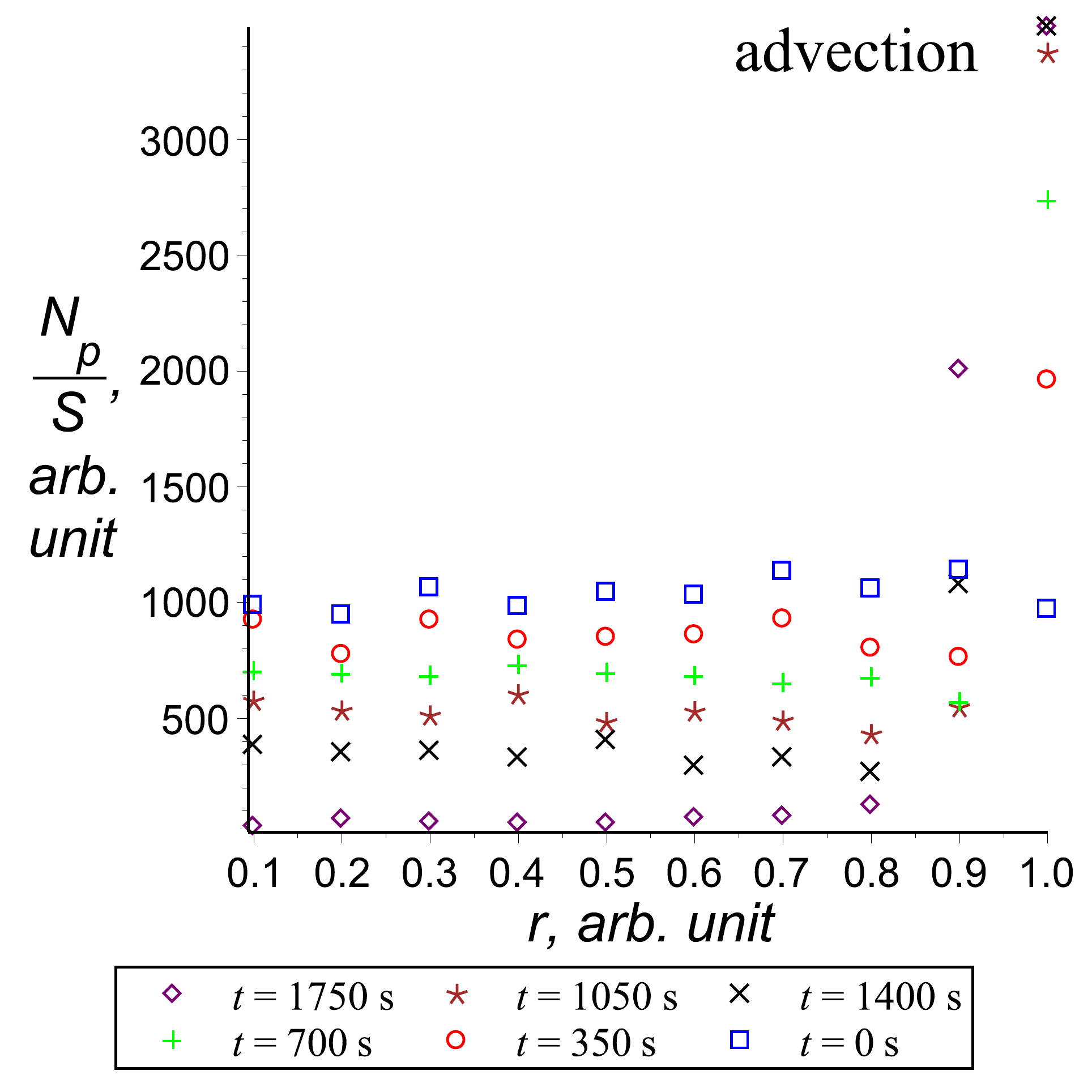}(b)
\end{minipage} 
\caption{Dynamics of colloidal particles in the case of advective transfer: (a) arrangement of particles at different times, (b) the density distribution of particles.}
\label{fig:advection_structures_and_densities}
\end{figure}

The calculation result of particle transfer by a single fluid flow is shown in the fig.~\ref{fig:advection_structures_and_densities}. We observe the formation of dendritic-like structures near the drop edge during the whole process  in this case (fig.~\ref{fig:advection_structures_and_densities}a). The particles do not mix randomly when diffusion is absence. This leads to the fact that the particle, bumping with another sticked particle in front, stops moving further to the fixation boundary. This sticked particle interferes with another particles behind the current one to move relative to the direction of flow. Their tree-like shape distribution is formed  in this way.
The density distribution of particles in the central part of the domain decreases, and their number near the three-phase boundary increases (fig.~\ref{fig:advection_structures_and_densities}b).

\begin{figure}[h]
\begin{minipage}{14pc}
\includegraphics[width=14pc]{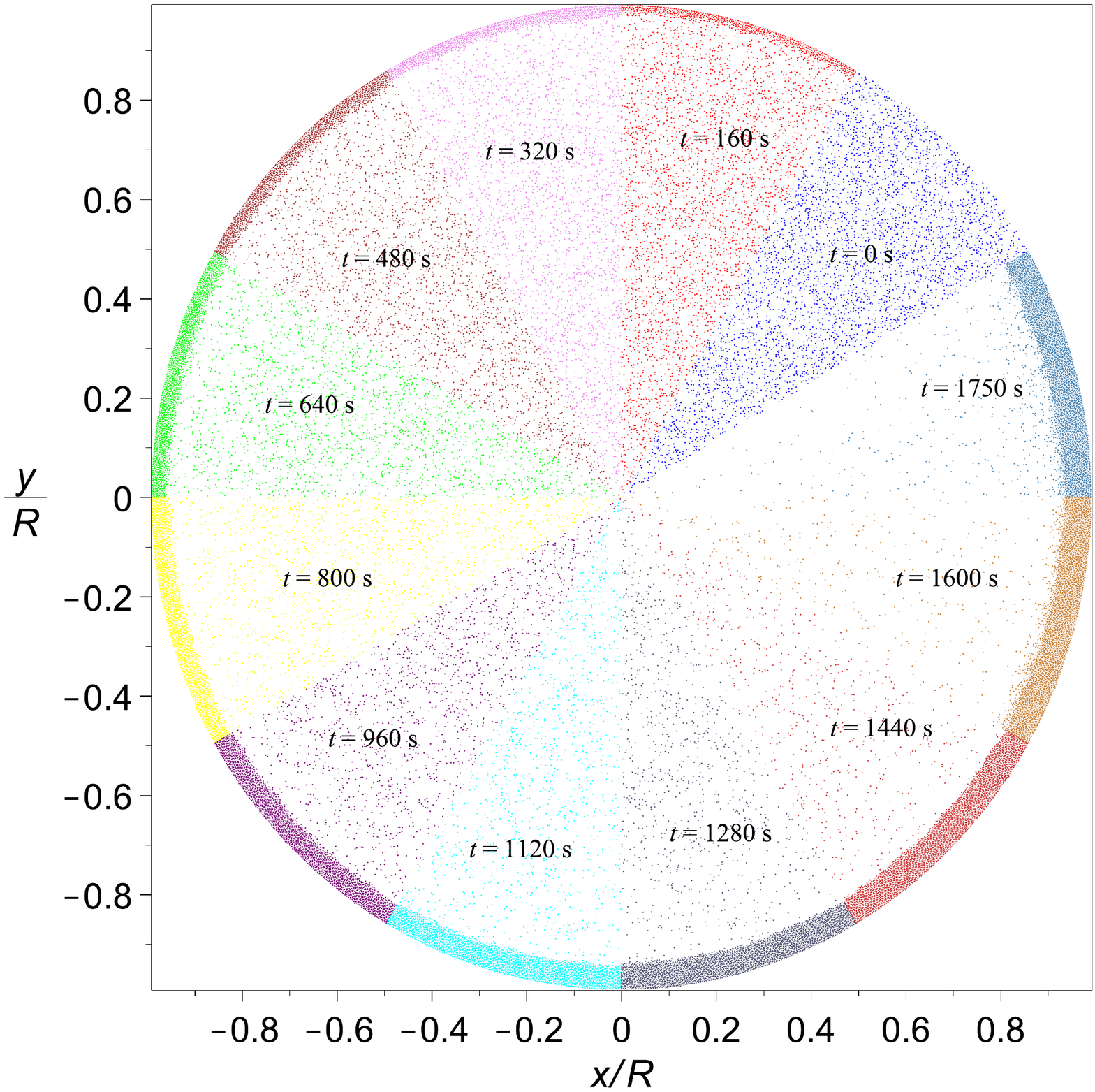}(a)
\end{minipage}\hspace{2pc}%
\begin{minipage}{14pc}
\includegraphics[width=14pc]{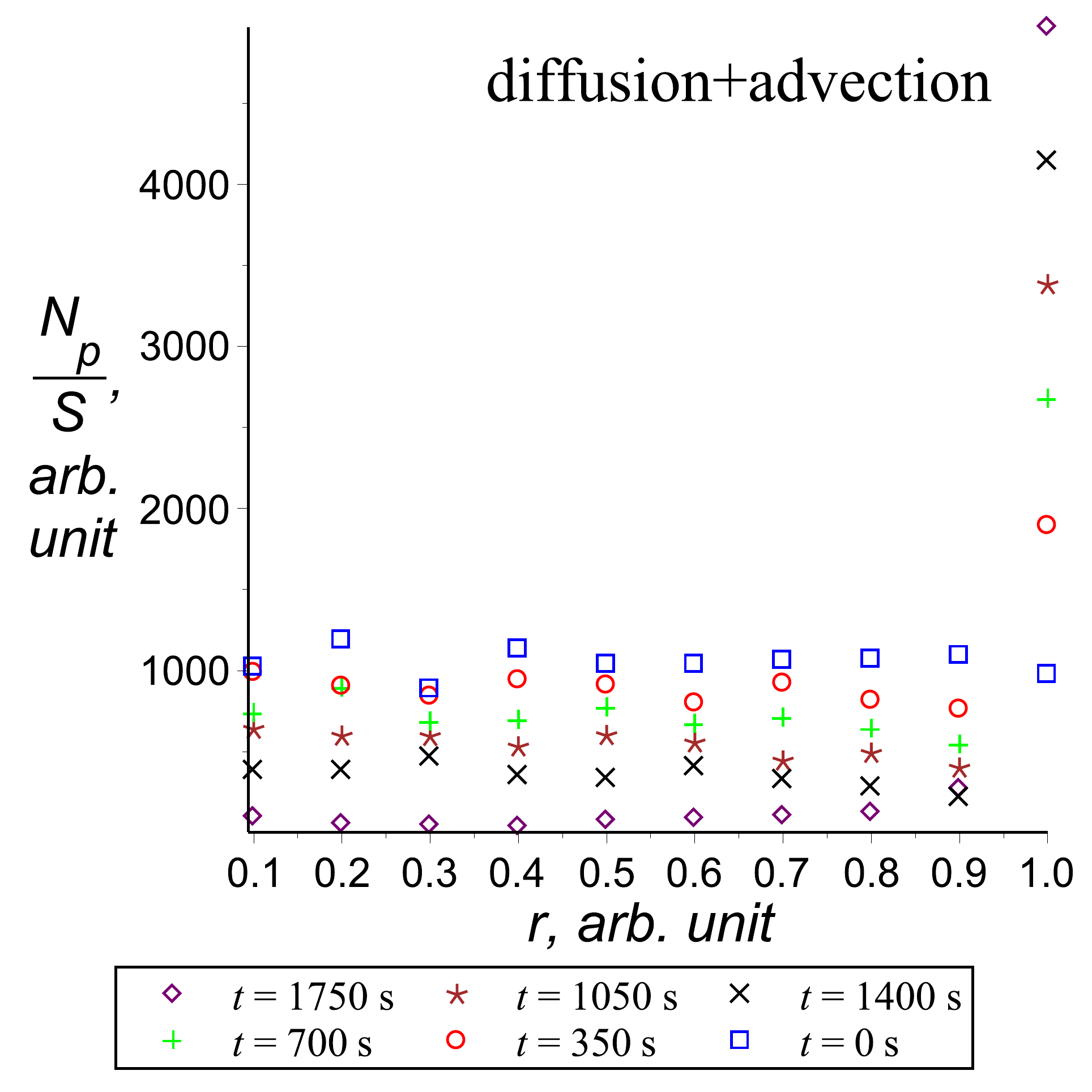}(b)
\end{minipage} 
\caption{Dynamics of colloidal particles in the case of advective and diffusion transfer: (a) arrangement of particles at different times, (b) the density distribution of particles.}
\label{fig:diffusion_advection_structures_and_densities}
\end{figure}

The numerical data for the third case are shown in fig.~\ref{fig:diffusion_advection_structures_and_densities}. Advection and diffusion leads to the formation of a more ordered ring deposition (fig.~\ref{fig:diffusion_advection_structures_and_densities}a). The number of particles near the contact boundary increases significantly over time (fig.~\ref{fig:diffusion_advection_structures_and_densities}b). These numerical results correspond qualitatively to the experimental observations of the coffee ring effect~\cite{Deegan2000}. But the change of close packing of particles (quasicrystal) to the inner amorphous part of the ring when moving from the edge of the drop towards its center is not observed here (fig.~\ref{fig:diffusion_advection_structures_and_densities}a). These simulation results differ from experimental data~\cite{Marin2011}.

\section{Conclusion}
Discrete (non-lattice) mathematical model describing the mass transfer of colloidal particles during the evaporation of a droplet on a substrate in the pinning mode is constructed. The concept of a fixation radius $ R_f $, which represents a moving boundary, in the region of which the thickness of the liquid layer is comparable with the particle size, is introduced. The Brownian motion of particles is modeled by the Monte Carlo method. The radial flow velocity of the liquid is calculated analytically. The cases are considered separately taking into account only diffusion, advection and both effects at the same time. The final particle distribution after the droplet drying for these cases is obtained. The resulting structure resembles a spot if there is no fluid flow and the particles move randomly. The same was observed in the experiment~\cite{Deegan2000}. The model predicts the formation of dendritic structures in the periphery region in the presence of only one advection in the system. Numerical calculations taking into account both factors (diffusion and advection) show the formation of a ring structure at the place of a dried drop, which corresponds to experimental observations~\cite{Deegan2000}. Unfortunately, it was not possible to mathematically describe the effect of the formation of a quasicrystal on the outer part of the ring and the amorphous structure in its inner zone on the basis of the proposed model. This makes one wonder whether diffusion and advection are the only determinants of this phenomenon, according to the authors of the experiment~\cite{Marin2011}. It is probably necessary to take into account additional effects that can also affect this process, for example, surface tension of the liquid~\cite{Kralchevsky1994}. Thus, this question remains open.

\ack The problem statement belongs to Yu.\,Yu.~Tarasevich. The author is grateful to L.\,Yu.~Barash, I.\,V.~Vodolazskaya, A.\,V.~Eserkepov and Yu.\,Yu.~Tarasevich for fruitful discussions and useful advice. This work was supported by the Russian Science Foundation (project 18-71-10061).

\section*{References}
\bibliographystyle{iopart-num}
\bibliography{references}

\end{document}